# Dipole-Spread Function Engineering for 6D Super-Resolution Microscopy

**Tingting Wu and Matthew D. Lew***

Department of Electrical and Systems Engineering, Washington University in St. Louis, MO, USA
Emails: Tingting Wu (wu.t@wustl.edu); Matthew D. Lew (mdlew@wustl.edu)

## Abstract

Fluorescent molecules are versatile nanoscale emitters that enable detailed observations of biophysical processes with nanoscale resolution. Because they are well-approximated as electric dipoles, imaging systems can be designed to visualize their 3D positions and 3D orientations, so-called dipole-spread function (DSF) engineering, for 6D super-resolution single-molecule orientation-localization microscopy (SMOLM). We review fundamental image-formation theory for fluorescent dipoles, as well as how phase and polarization modulation can be used to change the image of a dipole emitter produced by a microscope, called its DSF. We describe several methods for designing these modulations for optimum performance, as well as compare recently developed techniques, including the double-helix, tetrapod, crescent, and DeepSTORM3D learned point-spread functions (PSFs), in addition to the tri-spot, vortex, pixOL, raPol, CHIDO, and MVR DSFs. We also cover common imaging system designs and techniques for implementing engineered DSFs. Finally, we discuss recent biological applications of 6D SMOLM and future challenges for pushing the capabilities and utility of the technology.

## Introduction

To break the Abbé diffraction limit, which prevents optical microscopy from resolving neighboring emitters closer than $\sim\lambda/2\mathrm{NA}$,[1] where $\lambda$ is the optical wavelength and NA represents the numerical aperture of the objective lens, single-molecule localization microscopy (SMLM) utilizes fluorophores that blink over time. These molecules can be switched between dark and emissive states through transient binding of dyes to targets[2–4], photoactivation[5], photochemical switching,[6] and a variety of other mechanisms[7,8]. If the concentration of actively emitting fluorophores is sparse enough such that their images or point spread functions (PSFs) are well separated, then an appropriate SMLM image analysis algorithm or neural network can measure each molecule's position with high precision[9]. These blinking events can be accumulated over time into a super-resolved reconstruction of the biological target of interest (Fig. 1). Utilizing the standard diffraction-limited PSF, the 2D positions of individual emitters can be measured with high precision (typically ~15 nm with 1000 photons detected per localization and 10 background photons per pixel). It is well known that the standard PSF is poorly suited for localizing point-like emitters in 3D, and thus, many coded aperture methods have been developed for 3D SMLM.[10].

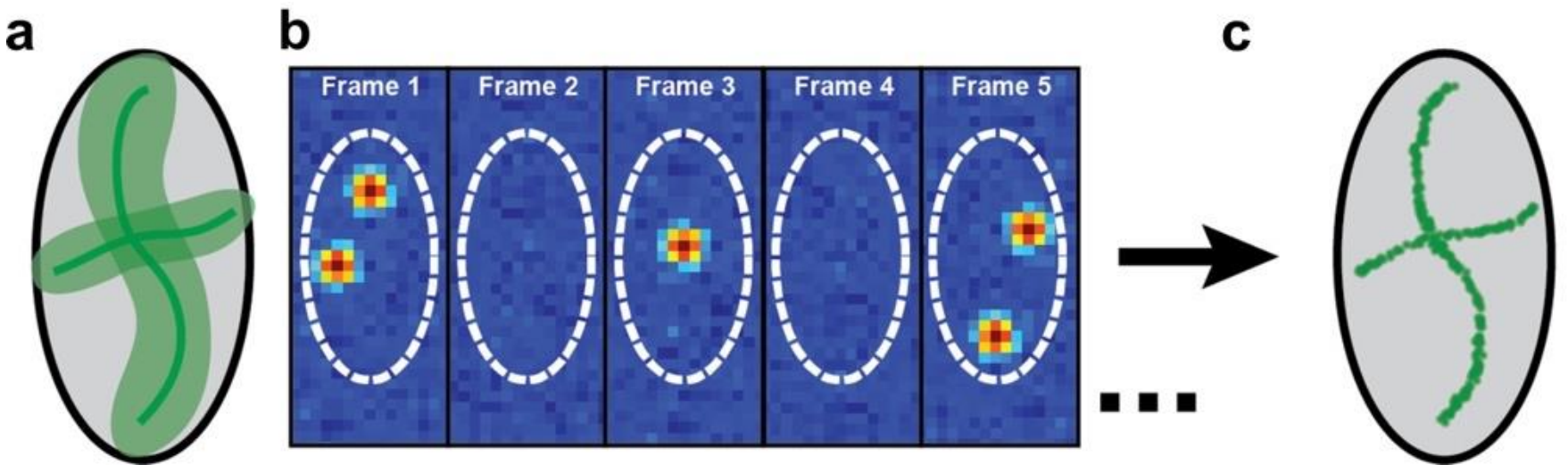

**Fig. 1** Super-resolved fluorescence imaging via single-molecule localization microscopy (SMLM). (a) Diffraction-limited image of the target. (b) Single-molecule images where the PSFs of individual emitters are well-separated spatially over time. (c) Super-resolution image reconstructed by estimating the 2D positions of each emitter shown in (b). Reprinted by permission from the American Chemical Society: *Chemical Reviews*, Three-Dimensional Localization of Single Molecules for Super-Resolution Imaging and Single-Particle Tracking, von Diezmann, L., et al., © 2017.

Here, we consider fluorescent molecules as oscillating electric dipoles (Fig. 2(a)), and thus, their radiation patterns, which contain characteristic intensity and polarization distributions (Fig. 2(b)), contain information about their orientations with respect to the imaging system. These orientations can, for example, probe the chemical environment surrounding the fluorescent molecule and provide additional insights into various biochemical processes[11–14]. Perhaps unsurprisingly, the standard PSF is also suboptimal for measuring the orientations of fluorescent molecules[15,16], so we review methods for measuring the 3D position and 3D orientation of fluorophores, thereby facilitating super-resolved single-molecule orientation-localization microscopy (SMOLM).

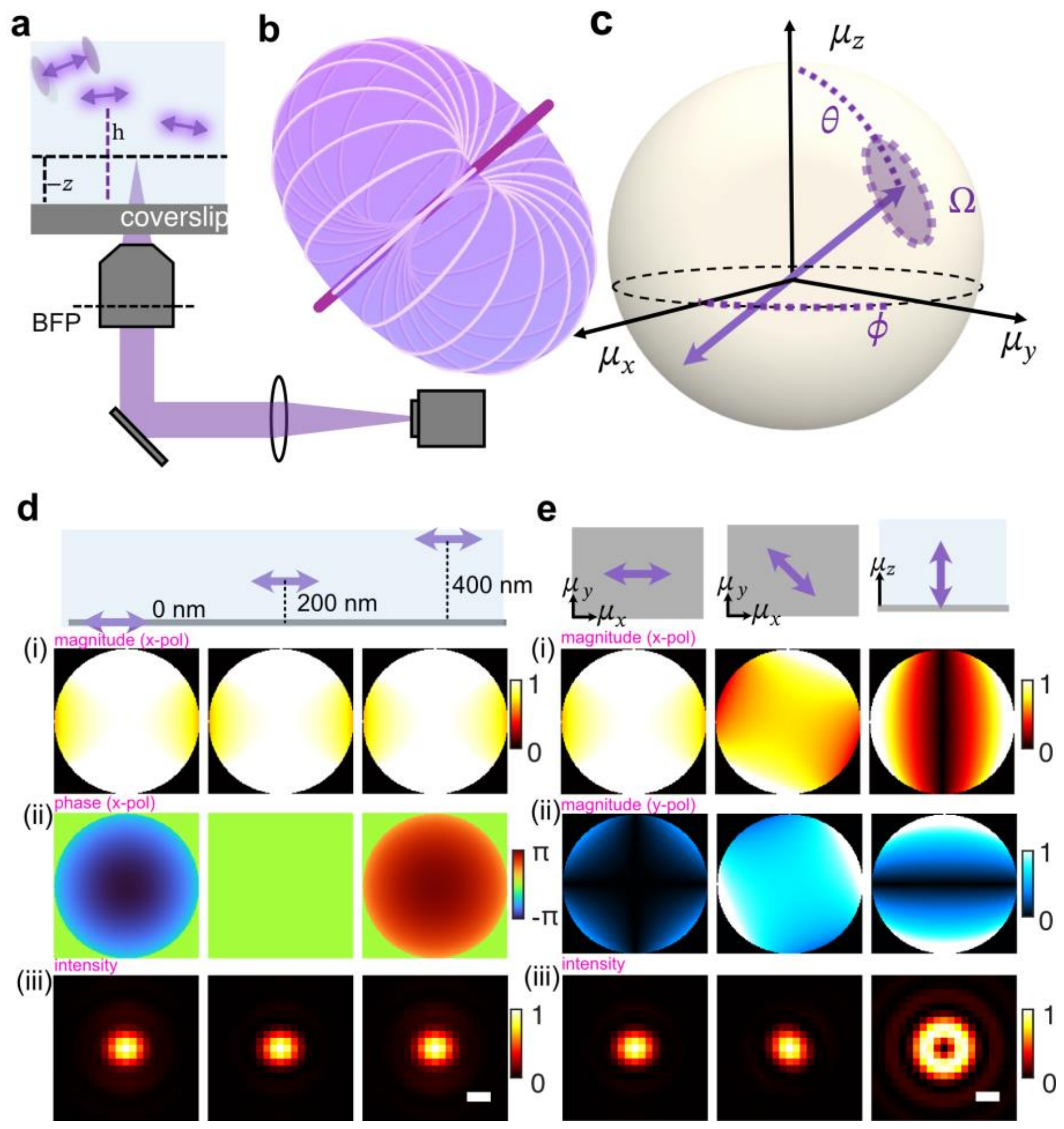

**Fig. 2** Imaging dipole-like emitters. (a) Schematic of a standard microscope with an emitter located at height h and an objective focused at a height -z. (b) The emission pattern (white lines) of an oscillating dipole. The dipole's emission probability (depicted as the distance separating the dipole and the purple surface) is proportional to $sin^2\,\eta$, where $\eta$ is the angle between the emission dipole moment and the direction of fluorescence emission. (c) Modeling the rotational "wobble" of a dipole within a cone oriented at $[\theta, \phi]$ with solid angle $\Omega$ . (d) (i) The magnitude and (ii) phase of the x-polarized electric field at the back focal plane (BFP), and (iii) the unpolarized intensity (x and y polarizations summed together) at the image plane for dipoles oriented along $\mu_x$. The three dipoles (left to right) are located at $h = 0$ nm, 200 nm, and 400 nm. (e) The magnitude of the (i) x-polarized and (ii) y-polarized electric field at the BFP, and (iii) the unpolarized intensity at the image plane for dipoles located at h=200 nm. The three dipoles (left to right) have orientations $[\theta, \phi, \Omega]$ of $[90°, 0°, 0\ sr]$, $[90°, 45°, 0\ sr]$, and $[0°, 0°, 0\ sr]$. For (d) and (e), the microscope is focused at $z = -200$ nm, and the refractive indexes of the immersion medium of the objective and the target sample are equal (n=1.518).

Examining the fluorescence collected from a single molecule at the back focal plane (BFP, and also called the pupil or Fourier plane) and the image plane of the microscope shows how imaging systems can be engineered for superior performance (Fig. 2(d,e)). Molecules with varying axial (z) positions (Fig. 2(d)) and orientations (Fig. 2(e)) clearly exhibit different intensity patterns at the BFP, but the intensity patterns in the image plane, called the molecule's dipole spread function (DSF), are extremely similar. Thus, the standard DSF has poor sensitivity for measuring the 3D positions and 3D orientations of single fluorophores. To solve this problem, many techniques modulate the phase and/or polarization of the fluorescence distribution at the BFP to better encode this 6D information into the DSFs measured in the image plane. We note that many other techniques modulate the polarization and/or spatial distribution of the illumination laser to measure molecular orientation; these techniques are outside the scope of this chapter, but refer interested readers to other literature[13,17–22].

## Forming images of dipoles: the dipole-spread function (DSF)

We model a fluorophore as a dipole-like emitter[23–25] with a mean orientation of $[\theta, \phi]$ in spherical coordinates, or equivalently a transition dipole moment $\boldsymbol{\mu} = [\mu_x, \mu_y, \mu_z]$ in Cartesian space, and a "wobble" solid angle Ω that characterizes its rotational diffusion[26] (Fig. 2(c)). Using vectorial Green's functions[27], the electric field $\boldsymbol{E}_{\text{BFP}}$ distribution at the BFP of the imaging system of a dipole with orientation $\boldsymbol{\mu}$ can be calculated as

$$\boldsymbol{E}_{\text{BFP}}(\boldsymbol{\mu}) = \begin{bmatrix} \boldsymbol{E}_{\text{BFP}}^{x} \\ \boldsymbol{E}_{\text{BFP}}^{y} \end{bmatrix} = \begin{bmatrix} \boldsymbol{g}_x^x, \boldsymbol{g}_y^x, \boldsymbol{g}_z^x \\ \boldsymbol{g}_x^y, \boldsymbol{g}_y^y, \boldsymbol{g}_z^y \end{bmatrix} \begin{bmatrix} \mu_x \\ \mu_y \\ \mu_z \end{bmatrix} \in \mathbb{C}^{2\times U\times V}, \tag{1}$$

where the BFP is sampled using $U \times V$ discrete spatial grid points, $\boldsymbol{E}_{\text{BFP}}^{q} \in \mathbb{C}^{U\times V}$ are the x and y polarized electric fields, respectively, and $\boldsymbol{g}_i^q \in \mathbb{C}^{U\times V}$ is a so-called "basis field" observed at the BFP from a dipole with orientation $\mu_i$. The superscript $q$ represents the two orthogonal polarizations (x and y) of the electric field, which may be detected separately in a polarization-sensitive imaging system or simply summed incoherently in a standard epifluorescence microscope. Since the light travels mostly parallel to the z direction in the paraxial limit due to the magnification of the objective lens, the electric field does not have a z-polarized component. Note that the polarization components may be expressed equivalently on any basis, e.g., the radial and azimuthal directions[28].

For a microscope focused on a nominal focal plane (NFP) of $-z$ (above the coverslip) and capturing fluorescence emitted by a dipole located at $h$ (also above the coverslip) (Fig. 2(a)), the electric field at the BFP has an additional defocus phase modulation given by

$$\boldsymbol{E}_{\text{BFP}}(\boldsymbol{\mu}, z, h) = \begin{bmatrix} \boldsymbol{p}_f \odot \boldsymbol{E}_{\text{BFP}}^x \\ \boldsymbol{p}_f \odot \boldsymbol{E}_{\text{BFP}}^y \end{bmatrix} \in \mathbb{C}^{2\times U\times V} \quad \text{and} \tag{2}$$

$$\text{with } \boldsymbol{p}_f = \exp\left(jk_z z\sqrt{1-(\boldsymbol{u}^2+\boldsymbol{v}^2)}\right)\exp\left(jk_h h\sqrt{1-\frac{n_z^2}{n_h^2}(\boldsymbol{u}^2+\boldsymbol{v}^2)}\right) \in \mathbb{C}^{U\times V}, \tag{3}$$

where $\odot$ is the element-wise multiplication operator, $\boldsymbol{u}, \boldsymbol{v} \in \mathbb{R}^{U\times V}$ are the locations of the $U \times V$ grid points within the BFP, $k_z = \frac{2\pi}{n_z}$ and $k_h = \frac{2\pi}{n_h}$ are the wavenumbers for light propagating in the immersion medium of the objective with a refractive index of $n_z$ (typically $n_z = 1.518$) and in the sample with a refractive index of $n_h$ ($n_h = 1.33$ for water), respectively, and $j = \sqrt{-1}$.

As shown in Eqns. 1 and 2, emitters with different orientations and axial positions have different electric field magnitude, phase, and polarization distributions at the BFP. Quantum estimation theory shows that measuring the 3D orientation and 3D position of a single molecule using this optical intensity at the BFP can achieve optimal measurement precision, namely, precision limited by the quantum Cramér-Rao bound[29–31]. However, the emission light from all fluorophores in the sample overlaps at the BFP; therefore, measuring molecular orientation at the BFP requires exciting and measuring one molecule at a time[32]. In contrast, well-separated PSFs at the image plane enable measuring the orientations and positions of multiple emitters simultaneously.

The tube lens of the microscope performs a Fourier transform $\mathcal{F}$ on the electric field $\boldsymbol{E}_{\text{BFP}}$ at the BFP to produce the electric field $\boldsymbol{E}_{\text{img}}$ at the image plane as

$$\boldsymbol{E}_{\text{img}} = \begin{bmatrix} \boldsymbol{E}_{\text{img}}^x \\ \boldsymbol{E}_{\text{img}}^y \end{bmatrix} = \begin{bmatrix} \mathcal{F}(\boldsymbol{E}_{\text{BFP}}^x) \\ \mathcal{F}(\boldsymbol{E}_{\text{BFP}}^y) \end{bmatrix}. \tag{4}$$

The intensities $\boldsymbol{I}_{\text{BFP}}$ at the BFP and $\boldsymbol{I}_{\text{img}}$ at the image plane are the squared magnitudes of the electric fields and therefore are given by

$$\boldsymbol{I}_{\text{BFP}} = \boldsymbol{E}_{\text{BFP}}\boldsymbol{E}_{\text{BFP}}^* \quad \text{and} \tag{5}$$

$$\boldsymbol{I}_{\text{img}} = \boldsymbol{E}_{\text{img}}\boldsymbol{E}_{\text{img}}^*, \tag{6}$$

where * represents the complex conjugate operator. Extending Eqns. 5 and 6 using Eqn. 1, the intensity distribution can be represented as a linear combination of six basis images weighted by the orientational second moments $\boldsymbol{m}$ as

$$\boldsymbol{I}_{\text{BFP}} = \left[\boldsymbol{B}_{xx}^{\text{BFP}}, \boldsymbol{B}_{yy}^{\text{BFP}}, \boldsymbol{B}_{zz}^{\text{BFP}}, \boldsymbol{B}_{xy}^{\text{BFP}}, \boldsymbol{B}_{xz}^{\text{BFP}}, \boldsymbol{B}_{yz}^{\text{BFP}}\right]\left[\langle\mu_x^2\rangle, \langle\mu_y^2\rangle, \langle\mu_z^2\rangle, \langle\mu_x\mu_y\rangle, \langle\mu_x\mu_z\rangle, \langle\mu_y\mu_z\rangle\right]^T = \boldsymbol{B}^{\text{BFP}}\left[m_{xx}, m_{yy}, m_{zz}, m_{xy}, m_{xz}, m_{yz}\right]^T = \boldsymbol{B}^{\text{BFP}}\boldsymbol{m}, \quad \text{and} \tag{7}$$

$$\boldsymbol{I}_{\text{img}} = \left[\boldsymbol{B}_{xx}^{\text{img}}, \boldsymbol{B}_{yy}^{\text{img}}, \boldsymbol{B}_{zz}^{\text{img}}, \boldsymbol{B}_{xy}^{\text{img}}, \boldsymbol{B}_{xz}^{\text{img}}, \boldsymbol{B}_{yz}^{\text{img}}\right]\left[\langle\mu_x^2\rangle, \langle\mu_y^2\rangle, \langle\mu_z^2\rangle, \langle\mu_x\mu_y\rangle, \langle\mu_x\mu_z\rangle, \langle\mu_y\mu_z\rangle\right]^T = \boldsymbol{B}^{\text{BFP}}\boldsymbol{m}, \tag{8}$$

where the basis matrices $\boldsymbol{B}^{\text{BFP}}$ at the BFP and $\boldsymbol{B}^{\text{img}}$ at the image plane are the system's responses to each orientational moment (Fig. 3) and are defined as

$$\boldsymbol{B}_{kl}^{\text{BFP}} = \begin{bmatrix} (\boldsymbol{p}_f \odot \boldsymbol{g}_{\text{k}}^x)(\boldsymbol{p}_f \odot \boldsymbol{g}_{\text{l}}^x)^* \\ (\boldsymbol{p}_f \odot \boldsymbol{g}_{\text{k}}^y)(\boldsymbol{p}_f \odot \boldsymbol{g}_{\text{l}}^y)^* \end{bmatrix} \quad \text{and} \tag{9}$$

$$\boldsymbol{B}_{kl}^{\text{img}} = \begin{bmatrix} \mathcal{F}(\boldsymbol{p}_f \odot \boldsymbol{g}_{\text{k}}^x)\, \mathcal{F}^*(\boldsymbol{p}_f \odot \boldsymbol{g}_{\text{l}}^x) \\ \mathcal{F}(\boldsymbol{p}_f \odot \boldsymbol{g}_{\text{k}}^y)\, \mathcal{F}^*(\boldsymbol{p}_f \odot \boldsymbol{g}_{\text{l}}^y) \end{bmatrix}. \tag{10}$$

The orientational second moments $m_{kl} = \langle\mu_k\mu_l\rangle$ are products of the first moments $\boldsymbol{\mu}(t)$ time-averaged over the exposure time $e$ as

$$\langle\mu_k\mu_l\rangle = \int_{t=0}^{t=e} \mu_k(t)\mu_l(t)\, dt, \tag{11}$$

where $k$ and $l$ represent the Cartesian components x, y, and z. Based on Eqns. 9 and 10, it is clear that any electric field modulation at the BFP will change system's basis images $\boldsymbol{B}^{\text{BFP}}$ at the BFP and $\boldsymbol{B}^{\text{img}}$ at the image plane. Similar to the basis electric fields $\boldsymbol{E}_{\text{BFP}} \in \mathbb{C}^{2\times U\times V}$, the basis matrices $\boldsymbol{B}^{\text{BFP}}$ and $\boldsymbol{B}^{\text{img}}$ contain both x- and y-polarized components (Fig. 3). Ding *et al*. and Zhang *et al*. have shown that directly imaging the two orthogonal polarization components separately and simultaneously enhance the precision of estimating molecular orientation[12,28].

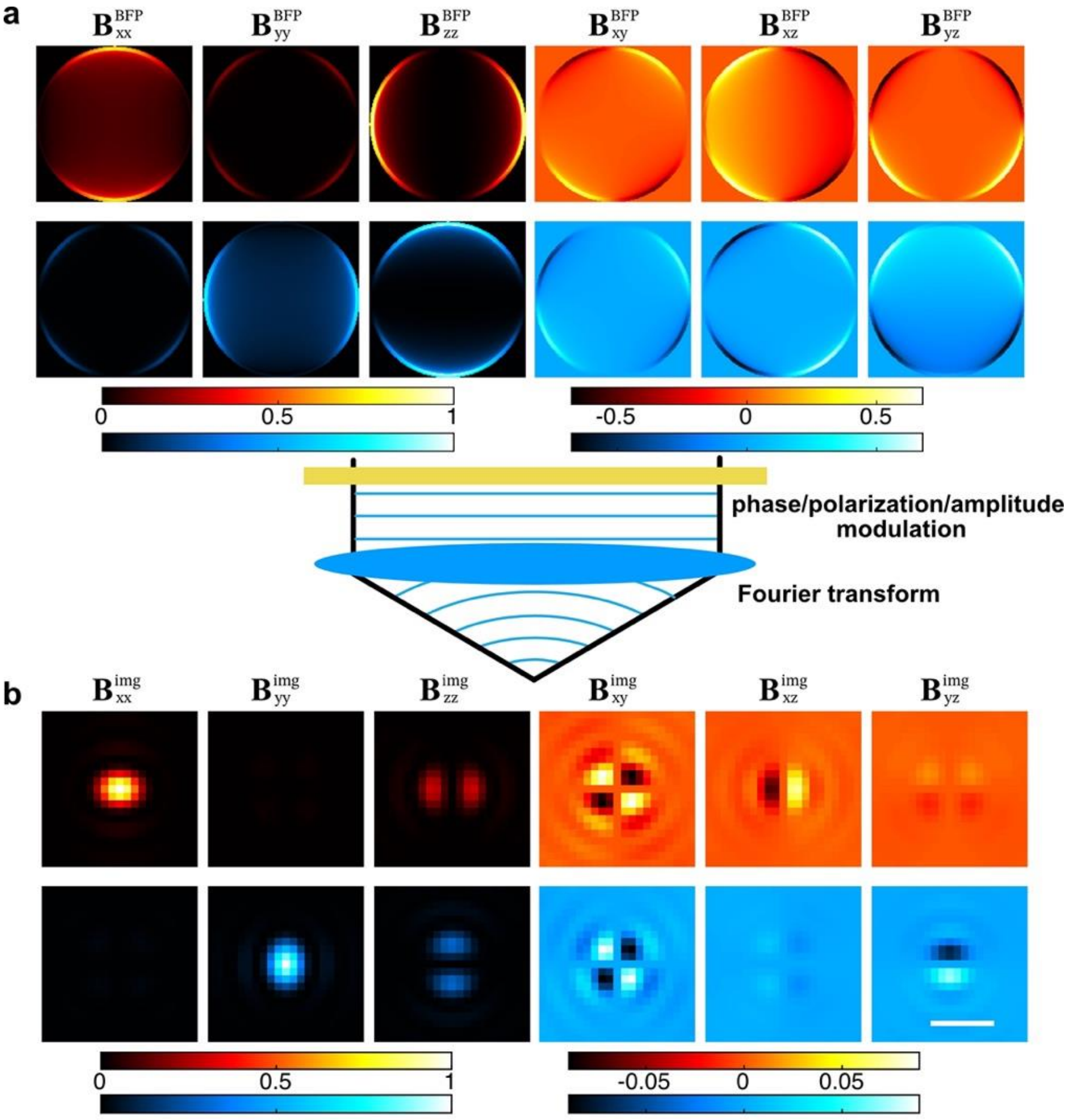

**Fig. 3** Basis images of x- and y-polarized microscopy. The (a) basis images at the back focal plane (BFP) are linked to the (b) basis images at the image plane through an optical Fourier transform $\mathcal{F}$ performed by the microscope tube lens on the corresponding electric fields. Top row (red): x-polarized light, bottom row (blue): y-polarized light. The electric field at the BFP can be modulated by phase, amplitude, and/or polarization masks; only the standard DSF (no modulation) is shown here. Scale bar: 500 nm.

One may model phase and polarization modulation at the BFP using a modulation tensor $\boldsymbol{J}$, given by

$$\boldsymbol{J} = \begin{bmatrix} \boldsymbol{J}_1 & \boldsymbol{J}_2 \\ \boldsymbol{J}_3 & \boldsymbol{J}_4 \end{bmatrix} = \begin{bmatrix} \boldsymbol{A}_1 \odot \exp(j\boldsymbol{P}_1) & \boldsymbol{A}_2 \odot \exp(j\boldsymbol{P}_2) \\ \boldsymbol{A}_3 \odot \exp(j\boldsymbol{P}_3) & \boldsymbol{A}_4 \odot \exp(j\boldsymbol{P}_4) \end{bmatrix}, \tag{12}$$

where $\boldsymbol{A}_m \in \mathbb{R}^{U\times V}$ and $\boldsymbol{P}_m \in \mathbb{R}^{U\times V}$ represent amplitude and phase modulation at BFP coordinates $(\boldsymbol{u}, \boldsymbol{v})$, respectively. Notably, the modulation $\boldsymbol{A}_m$, and $\boldsymbol{P}_m$ may vary with position. Since the number of detected photons before photobleaching is extremely limited in SMLM, one may maximize the signal to noise ratio by constraining $\boldsymbol{J}$ to be a unitary transformation such that

$$\boldsymbol{J} = e^{j\boldsymbol{\gamma}} \begin{bmatrix} \boldsymbol{\alpha} & -\boldsymbol{\beta}^* \\ \boldsymbol{\beta} & \boldsymbol{\alpha}^* \end{bmatrix}, \tag{13}$$

where $|\alpha_i|^2 + |\beta_i|^2 = 1$, $\gamma_i \in [0,2\pi]$, and the subscript $i$ refers to the $i^{\text{th}}$ element of $\boldsymbol{\alpha}, \boldsymbol{\beta} \in \mathbb{C}^{U\times V}$, $\boldsymbol{\gamma} \in \mathbb{R}^{U\times V}$, and $|\cdot|$ represents the magnitude of a complex number. We therefore obtain the constraints

$$|\boldsymbol{A}_1| = |\boldsymbol{A}_4| = \sqrt{1 - |\boldsymbol{A}_2|^2} = \sqrt{1 - |\boldsymbol{A}_3|^2} \quad \text{and} \tag{14}$$

$$\boldsymbol{P}_1 + \boldsymbol{P}_4 = \boldsymbol{P}_2 + \boldsymbol{P}_3 = 2\boldsymbol{\gamma}, \tag{15}$$

where $\boldsymbol{\gamma}$ can be treated as a universal phase modulation that is independent of polarization.

With modulation at the BFP, the x- and y-polarized electric fields and basis images become

$$\begin{bmatrix} \boldsymbol{g}_{k,\text{mod}}^x \\ \boldsymbol{g}_{k,\text{mod}}^y \end{bmatrix} = \begin{bmatrix} \boldsymbol{p}_f \odot \left(\boldsymbol{J}_1 \odot \boldsymbol{g}_k^x + \boldsymbol{J}_2 \odot \boldsymbol{g}_k^y\right) \\ \boldsymbol{p}_f \odot \left(\boldsymbol{J}_3 \odot \boldsymbol{g}_k^x + \boldsymbol{J}_4 \odot \boldsymbol{g}_k^y\right) \end{bmatrix} \tag{16}$$

$$\boldsymbol{B}_{kl}^{\text{img}} = \begin{bmatrix} \mathcal{F}(\boldsymbol{g}_{k,\text{mod}}^x)\mathcal{F}^*(\boldsymbol{g}_{l,\text{mod}}^x) \\ \mathcal{F}(\boldsymbol{g}_{k,\text{mod}}^y)\mathcal{F}^*(\boldsymbol{g}_{l,\text{mod}}^y) \end{bmatrix}. \tag{17}$$

Modulation at the BFP creates a shift-invariant DSF at the image plane. Recently, modulation outside of the Fourier plane has been explored. Due to the complex nature of shift-varying optical responses, such approaches require sophisticated algorithms to model light propagation and/or employ machine learning to design the appropriate modulation. Since these efforts are out of the scope of this review, we refer readers to reference [33].

## How to engineer a DSF

For estimating the 3D positions and 3D orientations of dipole-like emitters, imaging systems need 1) to avoid degeneracy (Fig. 4(a)), i.e., molecules with different 3D orientations or 3D positions need to generate distinct images on the camera; 2) to exhibit high precision, i.e., the dipole images must change dramatically if the molecule changes its orientation and/or position; and 3) have high detection sensitivity, i.e., images of dim dipoles must be easily detected above background noise.

Early methods for measuring orientation focused on tackling the degeneracy problem; e.g., simply measuring linear dichroism, i.e., the intensity ratio between two orthogonally polarized detection channels, cannot distinguish between fixed dipoles oriented ±45° relative to the polarization axes. Defocusing the microscope to create ring-shaped DSFs enable sensitive measurements of molecular orientation[34], and recently, more advanced image models and algorithms enable orientation measurements from focused images of immobile molecules[35,36]. Extending to 3D, the astigmatic[37] and double helix (Fig. 4(b))[38] PSFs break the symmetry of imaging emitters below and above the focal plane to enable 3D SMLM. On the other hand, emitters with different orientations show obvious differences in their intensity distributions at the BFP (Fig. 2(e)), and early designs divide the intensity at the BFP into multiple spots in the image plane for 3D orientation measurement, e.g., the bisected[39], quadrated pupil[40], and tri-spot (Fig. 5(a))[41] DSFs. While the double-helix PSF was not explicitly designed for orientation measurements, it can be used to make accurate 6D SMOLM images (Fig. 4(b))[42].

As improved iterative maximum likelihood estimators[43] and neural networks[44–46] for estimating orientation became available, new PSF designs aimed for optimal precision, which typically results in PSFs with complex shapes. To quantitatively compare how well these shapes encode orientation information, designers compute the Fisher information $K_x$ of a parameter $x$ contained within an image $\boldsymbol{I}$ as

$$K_x = \sum_{l=1}^{N} \frac{1}{I_l} \frac{\partial \boldsymbol{I}_l^T}{\partial \mathrm{x}} \frac{\partial \boldsymbol{I}_l}{\partial \mathrm{x}}, \tag{18}$$

where $l$ indexes the $N$ pixels within an image of a single molecule corrupted by Poisson noise and neglecting camera readout noise[47]. The inverse of Fisher information, called the Cramér-Rao (lower) bound (CRB)[47] quantifies the best-possible variance $\mathrm{Var}(\hat{x})$ of an unbiased estimator of parameter $x$, given by

$$\mathrm{Var}(\hat{x}) \geq {K_x}^{-1}. \quad (19)$$

By using the CRB as the loss function of an optimization algorithm, the tetrapod PSF (Fig. 4(c))[48,49] and pixOL DSF (Fig. 5(c))[50] minimize the CRB for estimating emitters' 3D positions and 3D orientations, respectively. One may also minimize the CRB of estimating distances between two emitters; this strategy produced the crescent PSF (Fig. 4(d)), a nearly optimal phase mask for estimating the 3D positions of closely spaced emitters[51].

A significant difficulty with this approach is optimizing degeneracy, precision, and detectability simultaneously, which often exhibit tradeoffs with one another. One natural solution is to design DSFs that directly maximize estimation performance on trial measurements using training data. In the context of 3D SMLM, Nehme *et al.* codesigned an estimation neural network and a phase mask to achieve high estimation precision and high detectability for images containing dense emitters[44] (Fig. 4(e)). Recently, a similar approach was used to design the arrowhead DSF[52]. For training purposes, the Jaccard index is typically used to quantify the success rate of the estimation algorithm.

DSFs can also be optimized for specific imaging conditions, e.g., molecules lying in the xy-plane or out-of-plane, dense concentrations of blinking molecules, high fluorescence background, or samples that cause relatively large optical aberrations. The duo-spot DSF[11] and radially and azimuthally polarized (raPol) DSF (Fig. 5(d))[28] have been shown to have high performance for emitters tilted out of the coverslip plane.

Many engineered PSFs have large footprints, especially those composed of multiple spots[40,41]. For samples containing dense emitters, images of neighboring molecules may suffer from severe overlap on the camera and become difficult or impossible to resolve. The vortex[53,54], pixOL[50], and CHIDO[55] DSFs all have small footprints, high precision, and high detectability (Fig. 5(b,c,e)). Other methods utilize polarization filters to accomplish similar goals, including 4polar-STORM[20] and POLCAM[22].

A key difficulty of measuring 3D positions and 3D orientations simultaneously is that the covariance between measurement parameters is often nonzero, leading to degraded estimation precision or, even worse, coupled biases in the estimates. To overcome this limitation, Zhang *et al.* designed a multi-view reflector microscope that separates fluorescence into radially and azimuthally polarized channels, and then further separates fluorescence at the pupil plane into 4 nonoverlapping regions on a camera[56] (Fig. 5(f)). In each channel, the DSFs are similar to that of a standard microscope, and each channel effectively views the sample from a slightly different direction and exhibits a smaller effective numerical aperture. A key feature of this design is that the position of a molecule, including its axial location, is encoded into the lateral position of the DSF in each channel. Orientation information, however, is contained within the relative brightness of the DSF in each channel, thereby decoupling position estimates from orientation estimates. The authors demonstrate robust, accurate, and precise 6D measurements of molecular position and orientation in the presence of refractive index mismatch.

## Implementing an engineered DSF

DSF engineering is most commonly achieved by placing optical elements at the BFP, as this position provides convenient access to Fourier space so that the light captured from each molecule can be manipulated identically and simultaneously. Single-molecule imaging typically uses an objective with a large numerical aperture (NA) to maximize the number of captured photons and improve imaging resolution. The physical pupil of many microscope objectives is typically located within the lens housing (Fig. 6(a)). Thus, to access the BFP, researchers use imaging relays[27], e.g., a pair of lenses forming a 4f system, to simultaneously form a conjugate BFP and a conjugate image plane (Fig. 6(b)). By adjusting the focal lengths of the two lenses, the physical size of the BFP can be adjusted to match the optical components used for DSF engineering.

Spatial light modulators (SLMs) are widely employed to modulate the phase, intensity, and/or polarization of the optical field. They are electronically programmable to provide pixel-level control, and two technologies are popular: micro-electro-mechanical (MEMS) deformable mirrors and liquid crystals (LCs). MEMS deformable mirrors use electromechanical actuators to deform a flexible mirror surface, thereby controlling phase locally near each actuator[57]. Alternatively, segmented MEMS mirrors feature multiple reflective surfaces whose tilts are controlled independently. Although MEMS deformable mirrors have fast modulation speeds (1-100 kHz) and are widely used in adaptive optics[58,59], they have relatively few independent mirror segments (typically hundreds to thousands) that exhibit a moderate degree of crosstalk and are thus less suitable for creating large-magnitude or discontinuous optical phase masks.

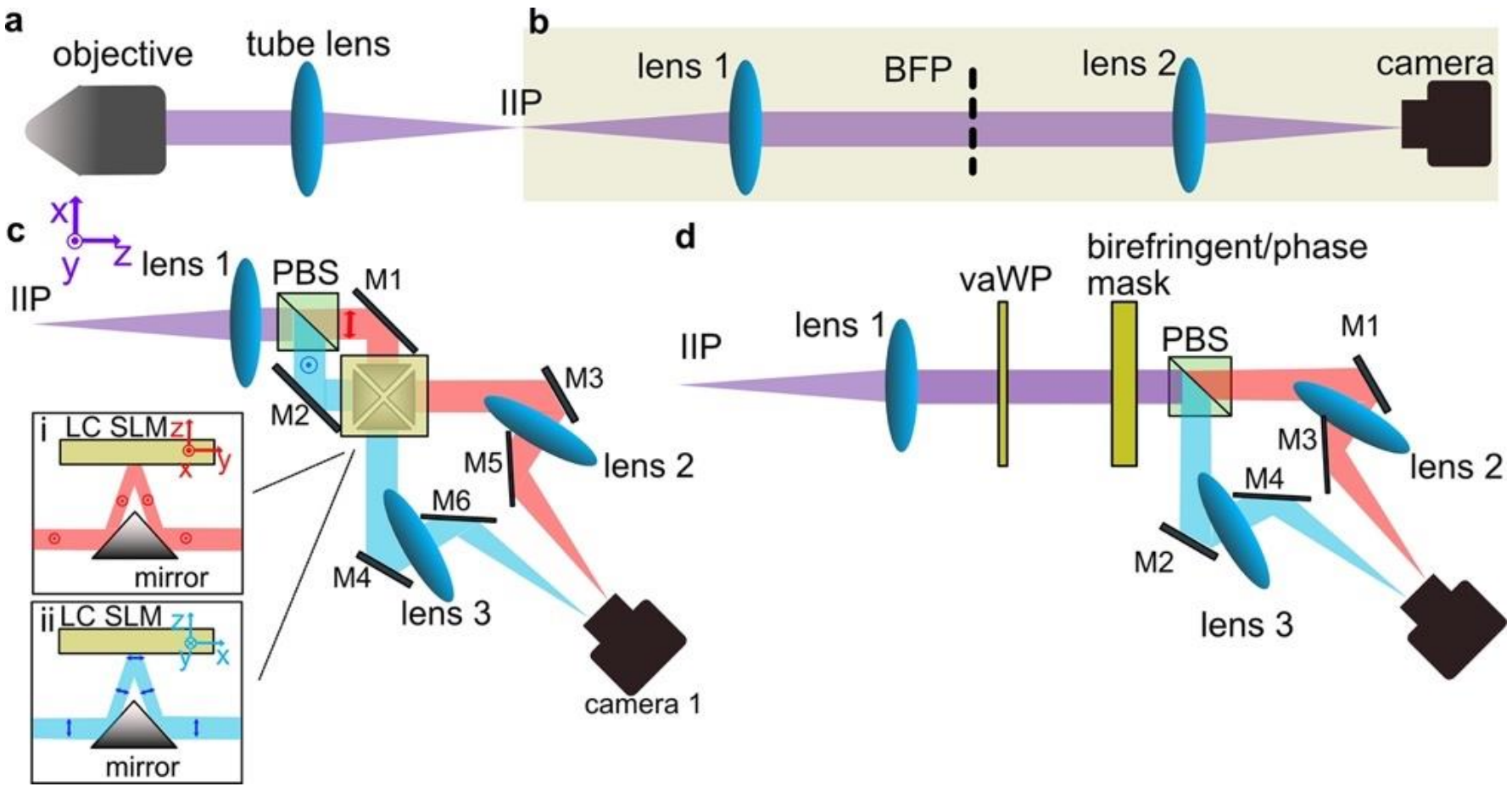

**Fig. 4** Imaging systems for implementing DSF engineering. (a) Fluorescence is collected by an objective. An intermediate imaging plane (IIP) is formed after the tube lens. (b) To access the back focal plane (BFP), a pair of lenses (1 and 2) is used to form a 4f system and focus the fluorescence onto a camera. (c) Modulating light using a reflective liquid crystal (LC) spatial light modulator

(SLM). The extended 4f system in (b) is reconfigured into (c) with the LC SLM placed at the BFP. A polarizing beam splitter (PBS) is added after the IIP to split the fluorescence into x- (red) and y-polarized (blue) channels. A pyramid mirror is used to reflect the light from two channels onto the surface of the reflective SLM. The pyramid mirror rotates x- and y-polarized light such that they are approximately parallel at the SLM surface (insets (i) and (ii)). Two lenses (2 and 3) focus the light onto two non-overlapping regions of a single camera. (d) Modulating light using a transmissive mask. For modulating polarization, a variable wave plate (vaWP) can be used to correct retardance variations. A birefringent or phase mask is placed at the BFP to modulate the light. A PBS splits the fluorescence into x- and y-polarized channels and two lenses (2 and 3) focus the light onto two non-overlapping regions of a single camera.

Alternatively, LC SLMs generally feature many more independent pixels ($10^4$ to $10^6$) but suffer from smaller fill factors and slower response times than MEMS deformable mirrors. Above a certain threshold of the applied electric field, the long axes of the molecules within an LC pixel will align with the field. Since these molecules are optically birefringent, any systematic change in molecular alignment will change the effective refractive index and, thus, the phase of the LC pixel for a suitable incident polarization. Since LC materials are dispersive, the user must calibrate the optical phase response as a function of the applied electric field before using them for DSF engineering. In addition, most LC SLMs cannot modulate two orthogonal polarization states simultaneously, and one must filter out the unwanted polarization or build a specific optical system to rotate and align the optical polarizations with the SLM's LCs (Fig. 6(c)).

Other than using programmable phase modulators, one may manufacture static phase masks that have higher photon efficiencies than SLMs. These masks are often manufactured using photolithography[60], but recent developments in 3D printing[61,62] have reduced the complexity of manufacturing high-quality masks. Double-helix[38], tetrapod[48,49], and vortex[53] phase masks are commercially available. Polarization optics, such as vortex half-waveplates, whose fast axes rotate systematically across the optic, have been useful for implementing radially and azimuthally polarized detection[28,56] in SMOLM.

More complex optical components, such as metasurface masks[63] and birefringent optical elements, can modulate the phases of the two orthogonal polarization states independently and simultaneously. The y-phi metasurface mask[64] uses a hexagonal array of elliptical nanoposts etched out of amorphous silicon to convert radially and azimuthally polarized light into x and y polarized light, respectively; Backlund *et al.* demonstrated this mask for removing dipole-induced localization bias in SMLM. CHIDO[55] uses a stress-engineered BK7 glass window[65] to encode position and orientation information into left- and right-handed circularly polarized detection channels of a fluorescence microscope. In fluorescence imaging, one must take care to compensate for polarization-dependent retardances stemming from any dichroic mirrors (DMs) in the optical path; a variable waveplate or birefringent polarization compensator can be used to restore the proper phase relationship between polarizations (Fig. 6(d)).

## Applications of SMOLM imaging

Three-dimensional SMLM reconstructs the 3D morphology of biological structures with nanoscale detail, but since it assumes fluorescent emitters are point sources, all vectorial information about molecular conformations and architecture is lost. DSF engineering inherently enhances imaging fidelity beyond standard 3D SMLM by enabling 6D imaging of 3D molecular orientations and 3D positions with nanoscale precision and accuracy. Here, we briefly review how the additional molecular orientation information provided by SMOLM yields rich biophysical insights beyond standard SMLM.

Tracking a molecule's position and orientation within soft matter is critical for understanding the intrinsically heterogeneous and complex interactions of its various components across length scales. Lu *et al.* measured the orientation of lipophilic probes within supported lipid bilayers (SLBs) with different chemical compositions[11]. SMOLM imaging showed that the orientational dynamics of the lipophilic probe Nile red (NR) are sensitive to cholesterol concentration; namely, as cholesterol increased, NR became increasingly parallel to the lipid molecules, and its rotational "wobble" decreased (Fig. 7(a)(i)). NR orientations could also be used to discern the degree of saturation of the lipid acyl chains (i.e., detecting the difference between DDPC, DOPC, and POPC) within the SLB (Fig. 7(a)(ii)). NR's binding behavior within a spherical SLB has been used to experimentally demonstrate accurate and precise 3D orientation and 3D position imaging using various DSFs [50] [22,56]. Zhang *et al.* track the positions and orientations of dyes in SLBs treated with cholesterol-loaded methyl-β-cyclodextrin (MβCD-chol)[28]. Before treatment, NR mostly exhibited small translational motions and large rotational movements. However, after MβCD-chol treatment, Nile red exhibited "jump diffusion" between regions of the intact membrane, i.e., translational motions were large but NR orientations were relatively fixed. Thus, SMOLM position and orientation information can complement one another to provide rich nanoscale detail about molecular interactions in biological systems.

Amyloid aggregates are signatures of various neurodegenerative disorders. Shaban *et al.*[66] and Ding *et al.*[12] showed that amyloidophilic dyes bind to amyloid fibrils in specific configurations. Both thioflavin T and Nile red bind parallel to the grooves formed by crossed-beta sheets within amyloid-beta fibers, i.e., along their long axes (Fig. 7(b)). However, Ding and Lew[54] also observed remarkable heterogeneity in NR orientations for other amyloid aggregates, thus implying similar variations in their nanoscale organization. Both ordered, "fibril-like" oligomers and disordered, amorphous oligomers were measured, even though these particles were aggregated under identical conditions and imaged in the same field of view. Similarly, Zhang *et. al*[56] observed the disruption of spherical SLBs by the aggregation of amyloid-beta. As the lipid membranes were infiltrated by amyloid-beta, NR exhibited increasingly varied orientations and wobble, which is in direct contrast to its well-ordered orientations and small wobble when in contact with amyloid fibers (Fig. 7(c)).

Dye orientations have also been utilized to characterize DNA structure. Several groups[13,18,43,53,67] showed that the intercalating dyes SYTOX Orange and YOYO-1 bind perpendicular to the long axis of the DNA (Fig. 7(d)). In contrast, the minor-grove binding dye, SiR-Hoechst, shows a broad orientation distribution when bound to DNA. Backer *et al.*[13,68] further used YOYO-1 to resolve the inclination of DNA base pairs within S-DNA as a function of DNA pulling geometry, torsional constraint, and negative supercoiling.

Furthermore, fluorophore orientations have been used to study the orientational order of actin stress fibers. Notably, the structure of each dye significantly affected how it was oriented relative to its parent fiber[67]; Alexa Fluor 488-phalloidin pointed along the fiber, Atto 633-phalloidin was oriented perpendicular to the fiber, and Alexa Fluor 647-phalloidin was relatively free to rotate and had no preferred direction. Rimoli *et al.*[20] thus used Alexa Fluor 488 to study actin stress fibers in U2OS and B16-F1 cells, noting that ventral fibers exhibit the highest alignment among the various types and that blebbistatin, a drug that inhibits myosin II activity, decreased this alignment slightly (Fig. 7(e)). Interestingly, lamellipodia SMOLM imaging suggests that actin filaments with preferred angular distributions and with more isotropic distributions in 3D coexist simultaneously.

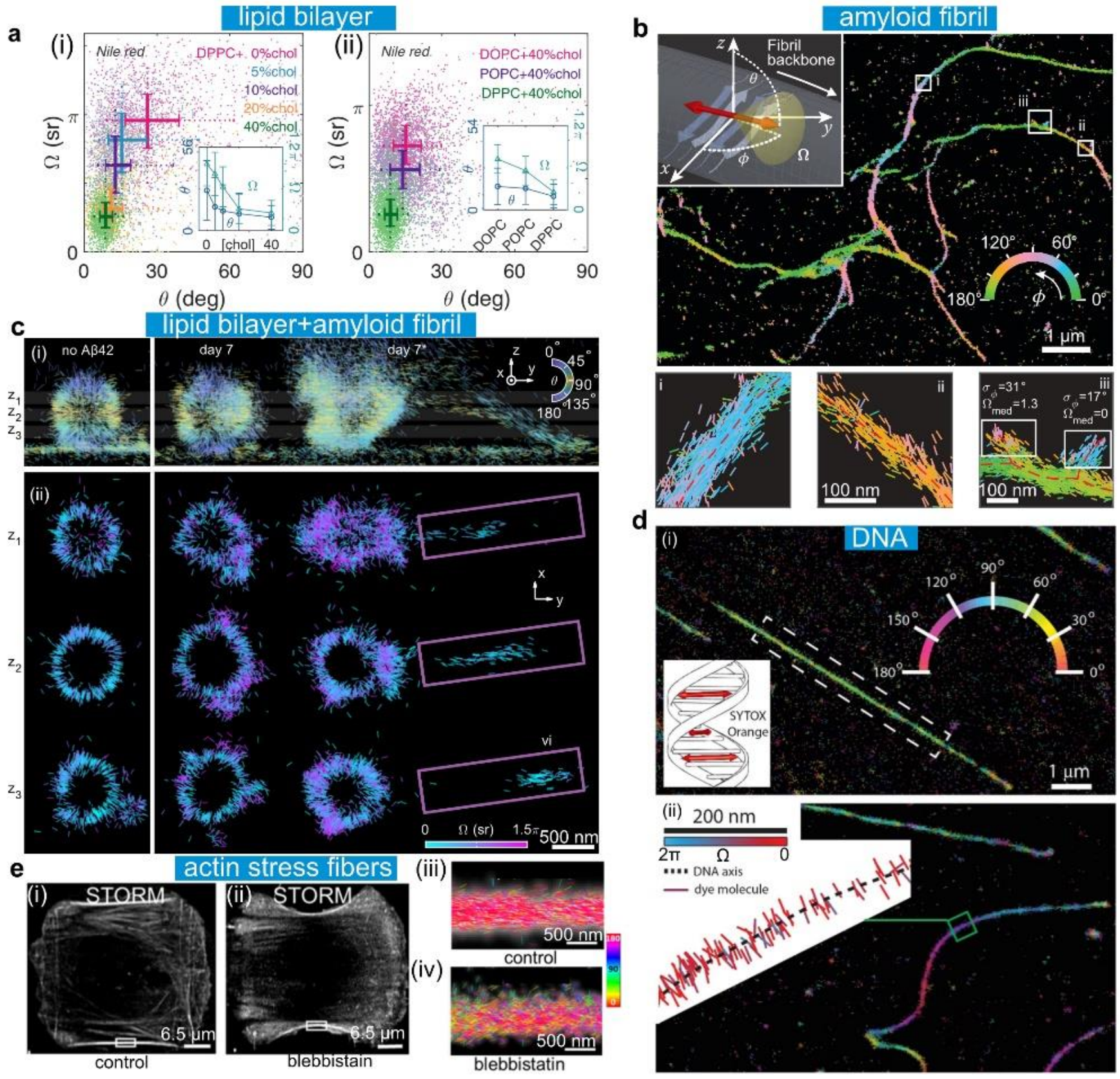

**Fig. 5** Applications of SMOLM imaging. (a) Measuring the tilt $\theta$ and wobble $\Omega$ of the lipophilic probe Nile red (NR) transiently bound to supported lipid bilayers (SLBs) (i) with various concentrations of cholesterol and (ii) composed of lipids whose acyl chains exhibit different degrees of saturation. (b) The azimuthal orientations $\phi$ of NR transiently bound to amyloid fibrils. Inset: NR molecules are mostly aligned parallel to the long axis of each fibril. (i,ii,iii) All NR orientation measurements within the boxes shown in (b). The lines are oriented and color-coded according to the estimates of $\phi$. (c) The orientations of NR transiently bound to spherical SLBs (350-nm radius) without amyloid beta (Aβ42) and after 7 days of incubation with Aβ42 monomers. (i) y-z and (ii) x-y views of the NR localizations at three axial slices through the SLB ($z_1 = 600, z_2 = 400, z_3 = 200$ nm). (d)(i)(ii) The orientation of SYTOX on λ-DNA strands, color-coded according to the mean azimuthal orientation $\phi$ of dye molecules measured within 30 nm voxels. (ii) inset: all SYTOX orientation measurements within the boxes shown in (ii). The lines are oriented according to the estimates of $\phi$ and color-coded with wobble angle $\Omega$. (e) STORM images of actin filament organization in (i) control and (ii) blebbistatin-treated U2OS cells. (iii-iv) The orientation of AlexaFluor 488 within the boxes shown in (i)(ii). The lines are oriented and color-coded according to the estimates of $\phi$. Only emitters with a small wobble angle are plotted here (refer to Ref.[20] for details). Panel (a) is reprinted by permission from the John Wiley and Sons: *Angewandte Chemie International Edition*, Single-Molecule 3D Orientation Imaging Reveals Nanoscale Compositional Heterogeneity in Lipid Membranes, Lu, J., et al[11], © 2020. Panel (b) is reprinted by permission from the Optica publication group: *Optica*, Single-molecule orientation localization microscopy for resolving structural heterogeneities between amyloid fibrils, Ding, T., *et al.*, [12] © 2020. Panel

(c) is reprinted by permission from the Springer Nature: *Nature photonics*, Six-dimensional single-molecule imaging with isotropic resolution using a multi-view reflector microscope, Zhang, O., *et al.* [56], © 2023. Panel (d) is reprinted by permission from the Optica publication group: *Optica*, Enhanced DNA imaging using super-resolution microscopy and simultaneous single-molecule orientation measurements, Backer, A. S., *et al.* [18], © 2016. Panel (e) is reprinted by permission from the Springer Nature: *Nature communications*, 4polar-STORM polarized super-resolution imaging of actin filament organization in cells, Rimoli, C. V., *et al.*[20], © 2022.

## Challenges and future opportunities

In this chapter, we described the theory, design, implementation, and applications of encoding information on the 3D positions and 3D orientations of single molecules into the images produced by a microscope, i.e., DSF engineering. These technologies have been demonstrated for *in vitro* studies of proteins and model biological systems, as well as cellular imaging. However, *in vivo* imaging of tissues and organisms requires additional development to design engineered DSFs that can cope with the scattering and autofluorescence within thick samples. One way to achieve this goal is to integrate engineered DSFs with advanced illumination strategies, e.g., light-sheet microscopy[69,70]. Furthermore, since some engineered DSFs encode the 6D information into subtle changes in DSF shape (Fig. 5), optical aberrations will influence and likely degrade estimation precision and accuracy. In this context, some combination of adaptive optics[71–73] and reliable methods to calibrate both phase and polarization aberrations will be critical for robust SMOLM imaging[74,75].

Decoding 6D information from DSFs in the presence of photon shot noise also requires further development in robust estimation algorithms that are unbiased, achieve precision close to the CRB, and are capable of processing data in real-time[22]. Machine learning will undoubtedly become more important in single-molecule image[44–46,76] and data analysis[77,78]. Further, most engineered DSFs require an additional 4f system and optical elements for modulating phase, polarization, or both. While these instruments can be built by trained optical engineers and microscopists, these designs are not suitable for the wider scientific community. Simpler, "plug and play" systems, such as using a polarization-sensitive camera[22] in place of a standard sensor, will aid in wider adoption of SMOLM techniques.

The utility of SMOLM depends upon linking the positions and orientations of fluorescent molecules to the biophysical processes of interest. Thus, dye-biomolecule interactions must be carefully characterized[11] or engineered to ensure that 6D SMOLM imaging data are relevant. In transient-binding[2,4,79,80] and DNA PAINT labeling schemes[81,82], the rigidity and orientational configuration of the binding moiety determine how the dye's orientational dynamics are coupled to the biomolecular target. For traditional covalent attachment of fluorophores to biomolecules, the structure and rigidity of the linker are critical[83,84]. While multi-functional attachment, i.e., linking the dye and biomolecule at more than one location, has been extraordinarily useful in a variety of single-molecule studies[85,86], more work is needed to make these schemes more facile and robust for a wider variety of biomolecular targets.

This chapter reviews DSF engineering for the specific use of 6D super-resolution microscopy. However, one can imagine controlling more degrees of freedom for each photon in the imaging system. For example, the wavelength of each photon can be encoded into the PSF to enable multi-color imaging using a grayscale camera[87]; ambient photons may be manipulated for object detection and depth estimation in computer vision[88]; and the wavelength and phase of each photon may be modulated such that an imaging system is optimized for measuring the separation between two objects, rather than their absolute positions[89]. We eagerly anticipate the new capabilities and discoveries that are enabled by higher dimensional control and imaging in next-generation coded optical imaging systems.